\documentclass{article}
\usepackage[T1]{fontenc} 
\usepackage[utf8]{inputenc} 
\usepackage{ismir,amsmath,cite,url}
\usepackage{graphicx}
\usepackage{booktabs}
\usepackage{amsfonts}
\usepackage{color}
\usepackage{soul}

\usepackage{lineno}

\title{Decoupling Magnitude and Phase Estimation with Deep R\lowercase{es}UN\lowercase{et} for Music Source Separation}

\oneauthor
{Qiuqiang Kong$^{1}$, Yin Cao$^{2}$, Haohe Liu$^{1}$, Keunwoo Choi$^{1}$, Yuxuan Wang$^{1}$}
{$^{1}$ByteDance $^{2}$ University of Surrey \\ $^{1}$\tt\small{\{kongqiuqiang, liuhaohe.7, keunwoo.choi, wangyuxuan.11\}@bytedance.com} \\
$^{2}$ \tt\small{yin.cao@surrey.ac.uk}}

\def\authorname{F. Author, S. Author, and T. Author}

\begin{document}

\maketitle
\begin{abstract}

Deep neural network based methods have been successfully applied to music source separation. They typically learn a mapping from a mixture spectrogram to a set of source spectrograms, all with magnitudes only. This approach has several limitations: 1) its incorrect phase reconstruction degrades the performance, 2) it limits the magnitude of masks between 0 and 1 while we observe that 22\% of time-frequency bins have ideal ratio mask values of over~1 in a popular dataset, MUSDB18, 3) its potential on very deep architectures is under-explored. Our proposed system is designed to overcome these. First, we propose to estimate phases by estimating complex ideal ratio masks (cIRMs) where we decouple the estimation of cIRMs into magnitude and phase estimations. Second, we extend the separation method to effectively allow the magnitude of the mask to be larger than~1. Finally, we propose a residual UNet architecture with up to 143 layers. Our proposed system achieves a state-of-the-art MSS result on the MUSDB18 dataset, especially, a SDR of 8.98~dB on vocals, outperforming the previous best performance of 7.24~dB. The source code is available at: \url{https://github.com/bytedance/music_source_separation}

\end{abstract}
\section{Introduction}\label{sec:introduction}
Music source separation (MSS) is a task to separate audio mixtures into individual sources such as vocals, drums, accompaniment, etc. MSS is an important topic for music information retrieval (MIR) since it can be used for several downstream MIR tasks including melody extraction \cite{salamon2014melody}, pitch estimation \cite{durrieu2011musically}, music transcription \cite{benetos2018automatic}, music remixing \cite{pons2016remixing}, and so on. MSS also has several direct applications such as Karaoke and music remixing. 

MSS methods can be categorized into signal processing based methods and neural network based methods. Several methods have been proposed for source separation such as non-negative matrix factorizations  (NMFs)~\cite{lee1999learning}. NMF decomposes a spectrogram into dictionaries and activations, and separated sources can be obtained by multiplying activations with different dictionaries. Sparse coding was used in \cite{plumbley2009sparse}, where audio signals are transformed into sparse representations for source separation. Independent component analysis (ICA) was used in \cite{davies2007source} by assuming that source signals are statistically independent. Other unsupervised source separation methods include modeling average harmonic structures in \cite{duan2008unsupervised}. Recently, neural network based methods became popular and have achieved state-of-the-art results in the MSS task. Those models include fully connected neural networks \cite{xu2014regression}, recurrent neural networks \cite{naithani2017low, uhlich2017improving}, convolutional neural networks \cite{jansson2017singing, chandna2017monoaural, stoter2019open, takahashi2018mmdenselstm, hennequin2020spleeter, liu2020channel, hu2020dccrn}, and time-domain separation models \cite{luo2019conv, lluis2018end, defossez2019music, stoller2018wave}.

First, several previously introduced MSS systems perform in the time-frequency domain and have achieved the state-of-the-art performance. However, many conventional spectrogram-based systems do not estimate the phases of separated sources \cite{jansson2017singing, chandna2017monoaural, stoter2019open, takahashi2018mmdenselstm, hennequin2020spleeter} and it upper bounds performance of MSS systems as we will show in this paper. Recently, several works were proposed to estimate the phases of clean sources. For example, PhaseNet \cite{takahashi2018phasenet}
treats the phase estimation as a phase classification problem, and PHASEN \cite{yin2020phasen} estimates the phase of clean sources using a separate neural network. Complex ideal ratio masks (cIRM) \cite{choi2018phase, wang2018supervised, tan2019complex} were also used for MSS. However, directly predicting the real and imaginary parts of cIRMs can be difficult, because the real and imaginary parts are sensitive to signal shifts in the time domain. In this paper, we propose to decouple the magnitude and phase for estimating cIRMs, which increases the performance of the source separation systems. We also elaborately design the magnitude estimation submodule to increase the upper bound of MSS systems.

Second, several magnitude or complex mask-based methods \cite{hu2020dccrn} usually limit the magnitude of masks to 1. Based on our analysis, this limits the upper bound of the performance of MSS systems. In this work, we observe that 22\% time-frequency bins in the cIRM have magnitudes larger than 1. To predict magnitudes with cIRMs larger than 1, we propose to combine the predictions of mask and spectrogram where the spectrogram term is a residual component to complement the mask prediction term. Therefore, we combine the advantage of mask and linear spectrogram based methods. All of mask magnitudes, and spectrograms and phases are learnt by a neural network. 

Third, we show that the previous UNets \cite{jansson2017singing, hennequin2020spleeter, hennequin2020spleeter, hu2020dccrn} with up to tens of layers have limited separation results in MSS. 
We show that the depth of neural networks are important for the MSS task. In this work, we propose a deep residual UNet with 143 layers. We propose using residual encoder blocks, residual intermediate layers, and residual decoder blocks to build the 143-layer residual UNet. We show that deep architectures significantly increase the MSS performance.

This paper is organized as follows. Section 2 introduces previous neural network based source separation systems and their limitations. Section 3 introduces our proposed system including the estimation of cIRMs and deep residual UNet. Section 4 shows experimental results and Section 5 concludes this work. 

\section{Backgrounds}\label{sec:background}

In this paper, we denote the time-domain signal of a mixture and a clean source as $ x \in \mathbb{R}^{L} $ and $ s \in \mathbb{R}^{L} $, respectively, where $ L $ is the number of samples of the signal. Their short-time Fourier transforms (STFTs) are denoted as 
$ X \in \mathbb{R}^{T \times F} $ and $ S \in \mathbb{R}^{T \times F} $, respectively. $ T $ and $ F $ correspond to the number of frames and frequency bins. Next, we describe several source separation methods.

\subsection{Approach 1: Direct Magnitude Prediction} \label{subsec:direct_pred}

In direct prediction approaches, a MSS system directly learns a mapping from $ |X| $ to $ |S| $, i.e., $\hat{|S|} = f(|X|)$.
Here, $ f $ can be any function approximator, such as a neural network of the fully connected, convolutional or recurrent types. 

Typically, a direct prediction method does not estimate the phases of separated sources. Instead, the phases of mixture is used to recover the STFT of separated sources:
\begin{equation} \label{mask}
\hat{S} = |\hat{S}| e^{j \angle X},
\end{equation}
\noindent where $ \angle X \in [-\pi, \pi]^{T \times F} $ is the phase of $ X $. Finally, we apply an inverse STFT $ \mathcal{F}^{-1} $ on $ \hat{S} $ to obtain the separated waveform $ \hat{s} = \mathcal{F}^{-1}(\hat{S}) $.

\subsection{Approach 2: Magnitude Mask}  \label{subsec:mag_mask}

In magnitude mask-based approaches, in order to perform source separation, a system predicts a mask, $ |\hat{M}| \in \mathbb{R}_{\geq 0}^{T \times F} $, that is applied to the input spectrogram element-wisely.

\begin{equation} \label{eq:mask}
\hat{|S|} = |\hat{M}| \odot |X|,
\end{equation}

The values of $\hat{M}$ can be continuous in the case of using ideal ratio masks (IRMs). The range of IRMs is often bounded between $[0, 1]$, assuming that the magnitudes of individual sources are smaller than the magnitudes of mixture. Furthermore, the magnitude is assumed to be either 0 or 1 in the case of ideal binary mask (IBM).

Similar to direct magnitude prediction, in magnitude mask-based methods, the phase of original mixture is used as an approximation of the phase of the separated sources.


\begin{figure}[t]
  \centering
  \centerline{\includegraphics[width=\columnwidth]{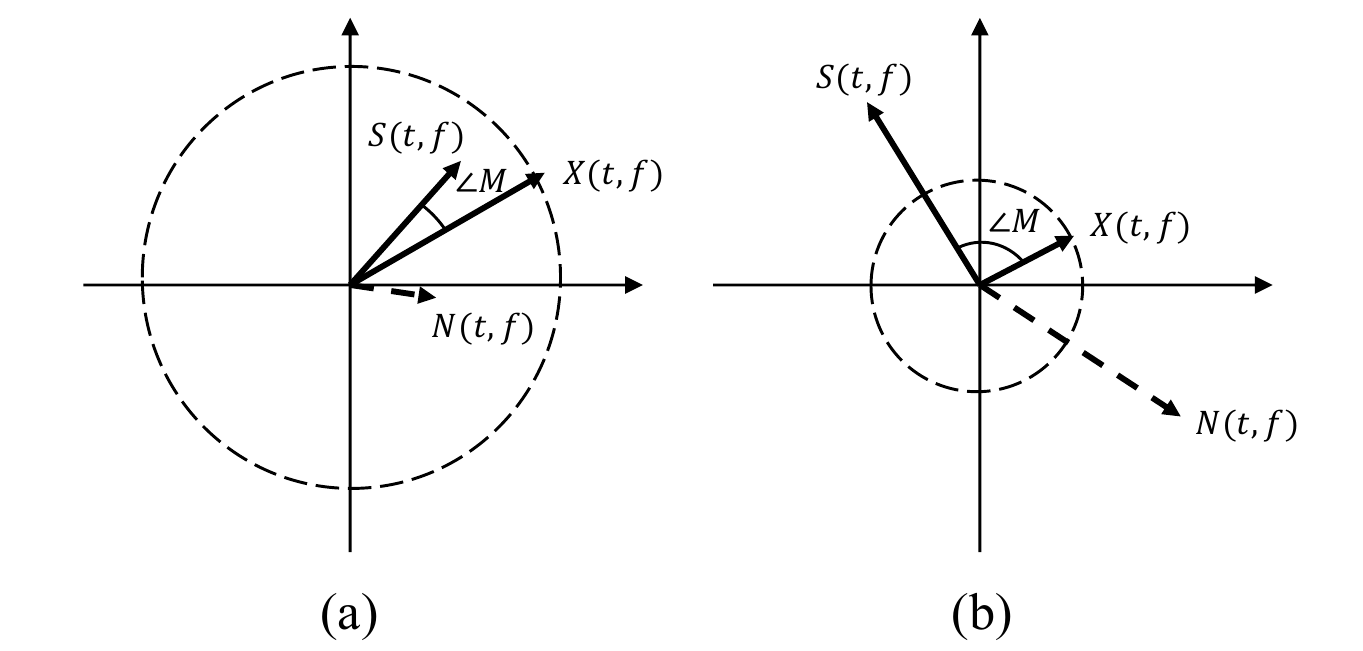}}
  \caption{Illustrations of a source signal $ s $, a noise $ n $, and mixture $ x $ on a complex plain. (a) is an example when $ |M(t, f)| $ smaller than 1 and (b) is an example when $ |M(t, f)| $ larger than 1.}
  \label{fig:cIRM}
\end{figure}

\begin{table*}[t]
\centering
\caption{The empirical upper bounds of MSS systems on MUSDB18. `acc.' indicates accompaniment. On the top row, numbers indicate the limit of the magnitude masks.}
\label{table:upper_bound}
\begin{tabular}{*{12}{c}}
 \toprule
 & Mixture & IBM & IRM (1) & IRM (inf) & cIRM (1) & cIRM (2) & cIRM (5) & cIRM (10) & cIRM (inf) \\
 \midrule
 vocals & -5.69 & 10.59 & 10.04 & 10.42 & 19.84 & 31.02 & 41.04 & 47.62  & 54.50 \\
 acc. & -5.68 & 16.10 & 15.31 & 15.97 & 26.54 & 37.62 & 47.33 & 53.51 & 60.63 \\
 bass & -6.36 & 7.17 & 6.05 & 6.07 & 17.99 & 27.88 & 37.86 & 44.30 & 54.12 \\
 drums & -4.30 & 8.75 & 8.03 & 8.61 & 19.10 & 30.38 & 39.91 & 46.45 & 56.08 \\
 other & -4.92 & 8.20 & 7.28 & 7.37 & 18.97 & 28.91 & 39.08 & 45.64 & 56.00 \\
 \bottomrule
\end{tabular}
\end{table*}

\begin{figure*}
 \centerline{
 \includegraphics[width=\textwidth]{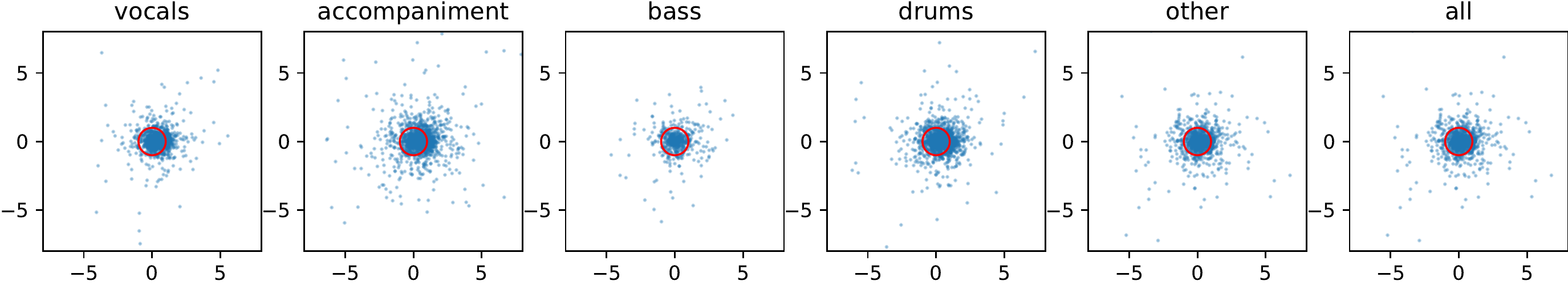}}
 \caption{cIRMs of vocals, accompaniment, bass, drums, other, and all sources, on the complex 2D plain. Unit circles are drawn in red.}
 \label{fig:cIRMs}
\end{figure*}

\subsection{Approach 3: Complex Mask}\label{subsec:comp_mask}

Accurate phase estimation becomes critical as the performance of the systems in Section~\ref{subsec:direct_pred} and \ref{subsec:mag_mask} has improved. 
Because of that, several works were proposed recently to take the phase estimation into consideration in the model. One ambitious approach is to directly predicting the complex STFT, as an extension of the direct magnitude prediction towards phase \cite{tan2019complex}.
However, accurate prediction of a complex STFT is challenging because the estimation of real and imaginary parts of a complex STFT is more difficult than the estimation of the magnitude. 
As an alternative, many methods have been introduced to predict complex masks of mixture STFT~\cite{wang2018supervised, tan2019complex, choi2018phase}.
In PhaseNet \cite{takahashi2018phasenet} and PHASEN \cite{yin2020phasen}, the authors proposed to predict the phases of signals independently from the magnitudes.

\subsection{Out-of-Phase and Masks }

In this work, we adopt cIRM-based methods for source separation due to their superior performance in phase estimation. A complex mask $ M \in \mathbb{C}^{T \times F} $ is calculated by:
\begin{equation} \label{eq:cIRM}
\begin{split}
M & = S / X \\
& = \frac{S_{r} + i S_{i}}{X_{r} + i X_{i}} \\
& = \frac{S_{r}X_{r} + S_{i}X_{i} + i(S_{i}X_{r} - S_{r}X_{i})}{X_{r}^{2} + X_{i}^{2}},
\end{split}
\end{equation}
\noindent where $ X_{r} $, $ S_{r} $ are real parts of $ X $ and $ S $ respectively, and $ X_{i} $, $ S_{i} $ are imaginary parts of $ X $ and $ S $ respectively. The perfect separation of $ S $ from $ X $ can be obtained by:
\begin{equation} \label{eq:mask_mul}
\begin{split}
S & = MX \\
& = |M||X|e^{j (\angle M + \angle X)}.
\end{split}
\end{equation}

\noindent Equation (\ref{eq:mask_mul}) shows that the separation of $ S $ from $ X $ includes a magnitude scaling and a phase rotation operation. The magnitude of cIRM ($ |M| $) controls how much the magnitude of $ X $ should be scaled, and the angle of cIRM ($ \angle M $) controls how much the angle of $ X $ should be rotated.

We now introduce an additive noise model, i.e., $ X = S + N $, which is illustrated in Fig. \ref{fig:cIRM}.
Here, we focus on each time-frequency bin of STFTs of source, noise, and mixture ($ S(t, f) $, $ N(t, f) $, and $ X(t,f) $, respectively).
Fig. \ref{fig:cIRM} (a) shows an example where the magnitude of cIRM $ |M(t, f)| $ is smaller than 1. This is modelled well in the existing methods where the ranges of complex mask is bound to $[0, 1]$. However, as illustrated in Fig.~\ref{fig:cIRM}~(b), $ |M(t, f)| $ can be larger than 1. As in the figure, this may happen when $S(t, f)$ and $N(t, f)$ are out of phase, since that makes the magnitude of mixture to be smaller than that of (individual) signal. 

\subsection{Empirical analysis of the effect of bounded magnitude mask}\label{subsec:empirical_analysis}

In this section, we empirically investigate the upper bound of the performance when the magnitude mask is bounded to be $< 1$, the common assumption in many previous methods. We use signal-to-distortion ratio (SDR) \cite{vincent2006performance} as an evaluation metric, which is defined as follow:

\begin{equation} \label{eq:sdr}
SDR(s, \hat{s}) = 10 \text{log}_{10} \frac{|| s ||^{2}}{|| \hat{s} - s ||^{2}}.
\end{equation}

\noindent A higher SDR indicates better separation results, and vice versa. Ideally, a perfect separation will lead to infinite SDR. We evaluate the upper bound of systems on the vocals, accompaniment, bass, drums and other instruments from the MUSDB18 dataset \cite{rafii2017musdb18}. 

The first column of Table \ref{table:upper_bound} shows the SDRs of using the mixture without separation as separated sources. The second column (IBM) shows the upper bound of the performance when IBMs are used. According to the third column (IRM (1)), using IRMs whose magnitudes are bounded in [0, 1] has slightly lower upper bounds compared to those of IBM. IRM uses the phases of mixture but not the phases of clean sources for separating sources so the upper bound SDR is limited. Unbounded IRM (IRM (0, inf), the fourth column) shows a small improvement over bounded IRM, but not significantly. 

Compared to IBM, IRM (1) and IRM (inf), the five cIRM columns show that the upper bounds are significantly higher when correct phase information is used. The upper bounds of cIRM (1) is higher than IRM (1) by around 10~dB. The improvement within cIRMs is also dramatic -- only by increasing the limit from cIRM (1) to cIRM (2), the upper bounds increase by more than 9.89~dB for all the instruments. When magnitude mask is unbounded, the SDR of cIRM (inf) is infinite in theory. Considering the numerical stability when calculating SDRs, we add a small $ \epsilon $ to the denominator of (\ref{eq:sdr}). We observed the SDRs of cIRM (inf) are higher than 50~dB for all the instruments.

\subsection{Distribution of cIRMs}
In this section, we visualize the distribution of cIRMs to show that there are much space to improve previous MSS systems. Fig. \ref{fig:cIRMs} shows the cIRMs of vocals, accompaniment, bass, drums, other and all sources. The horizontal and vertical axes show the real and imaginary parts of cIRMs calculated by (\ref{eq:cIRM}), where each point in Fig. \ref{fig:cIRMs} corresponds to a $ M(t,f) $. The unit circle shown in Fig. \ref{fig:cIRMs} corresponds to masks with magnitude values equal to 1. Fig \ref{fig:cIRMs} from left to right shows the cIRM distribution of vocals, accompaniment, bass, drums, other instruments, and all sources. It can be seen that there are many cIRMs that having magnitudes larger than 1. The ratio of cIRMs have magnitudes larger than 1 for vocals, accompaniment, bass, drums and others are 20.3\%, 34.5\%, 6.1\%, 26.9\% and 13.9\% respectively. Along with the analysis in Section~\ref{subsec:empirical_analysis}, this observation motivates our work to extend the bounded mask estimation methods to unbound mask estimation methods. 

Fig. \ref{fig:cIRMs} also shows that, the phases of cIRMs distribute evenly in all directions. However, spectrogram-based methods assume that the phases of cIRMs are all 0. This observation further justifies to predict the phases in a MSS system.


\section{Proposed System}
In this section, we propose a MSS system that incorporates phase estimation that is based on the proposed decoupling of magnitude and phase (Section~\ref{section:decoupling}). Furthermore, to overcome the limit of bounded magnitude mask as discussed in Section~\ref{sec:background}, we propose a modification to extend the mask estimation method that allows the magnitude of the resulting mask be larger than~1 (Section~\ref{section:attention}). Finally, we propose a deep Residual UNet with 143 layers, which is the first MSS architectures that is deeper than a hundred layers (Section~\ref{subsec:verydeep}). All the proposed systems are trained with a L1-loss that is computed on the waveform domain as illustrated in Figure~\ref{fig:resunet}.

\subsection{Decoupling Magnitude and Phase for cIRM Estimation}\label{section:decoupling}
Unlike previous works that directly predict real and imaginary parts of masks \cite{choi2018phase, hu2020dccrn}, we propose to decouple the magnitude and phase estimation for MSS so that we can 
optimize their designs separately.
We denote the complex mask to estimate as $ \hat{M} \in \mathbb{C}^{T \times F} $. As a part of the solution, our system outputs a bounded magnitude mask $ \hat{M}_{\text{mag}} \in \mathbb{R}^{T \times F} $ whose value is in $[0, 1]$. In practice, it is implemented by applying sigmoid function. 
Our system also outputs two more tensors, $ \hat{P}_{\text{r}} \in \mathbb{R}^{T \times F} $ and $ \hat{P}_{\text{i}} \in \mathbb{R}^{T \times F} $. Here, $ \hat{P}_{\text{r}} $ and $ \hat{P}_{\text{i}} $ are real and imaginary parts of $ \hat{M} $, respectively. Then, instead of calculating the angle $ \angle \hat{M} $ directly, we calculate its cosine value $cos \angle \hat{M}$ and sine value $sin \angle \hat{M}$ using $\hat{P}_{\text{r}}$ and $\hat{P}_{\text{i}}$ as follows:
\begin{equation} \label{eq:cos_sin}
\begin{split}
cos \angle \hat{M} & = \hat{P_{\text{r}}} / \sqrt{\hat{P_{\text{r}}}^2 + \hat{P_{\text{i}}}^2} \\
sin \angle \hat{M} & = \hat{P_{\text{i}}} / \sqrt{\hat{P_{\text{r}}}^2 + \hat{P_{\text{i}}}^2}. \\
\end{split}
\end{equation}
\noindent 
Then, we estimate the real and imaginary parts of cIRM by:
\begin{equation} \label{eq:recover_real_imag}
\begin{split}
\hat{M}_{\text{r}} = \hat{M}_{\text{mag}} cos \angle \hat{M} \\
\hat{M}_{\text{i}} = \hat{M}_{\text{mag}} sin \angle \hat{M} \\
\end{split}
\end{equation}
\noindent The cIRM $ \hat{M} = \hat{M}_{r} + j \hat{M}_{i} $ is a complex tensor, and is used to separate a target source from $ X $ by (\ref{eq:mask_mul}) which involves a magnitude scaling and a phase rotation operation. Finally, we apply an inverse STFT to obtain the separated waveform. 

\subsection{Combination of Bounded Mask Estimation and Direct Magnitude Prediction}\label{section:attention}
In previous works we show that directly predicting the unbound linear magnitude $ |\hat{S}| $ lead to the underperformance of the source separation system. To overcome the limit of the performance discussed in Section~\ref{sec:background}, we propose to combine a bounded mask and direct magnitude prediction to estimate the magnitude of cIRMs. 
The motivation is to use direct magnitude prediction as \textit{residual components}, one that complements the bounded magnitude mask. This is implemented as follow:

\begin{equation} \label{eq:attention}
|\hat{S}| = \text{relu}(\hat{M}_{\text{mag}} \odot |X| + \hat{Q})
\end{equation}

\noindent where 
$ \hat{Q} \in \mathbb{R}^{T \times F} $ is the direct magnitude prediction. In this way, we take the advantages of both of the methods. The ReLU operation ensures that the predicted magnitude is always larger than 0. The estimation of phase $ \angle \hat{M} $ by using $ \hat{P}_{\text{r}} $ and $ \hat{P}_{\text{i}} $ are the same as the one in Section \ref{section:decoupling}. Then, the separated STFT can be obtained by:
\begin{equation} \label{eq:attention_phase}
\hat{S} = |\hat{S}| e^{j (\angle \hat{M} + \angle X)},
\end{equation}
\noindent where $ |\hat{S}| $ is calculated by Eq.~ (\ref{eq:attention}). 

\begin{figure}
 \centerline{
 \includegraphics[width=\columnwidth]{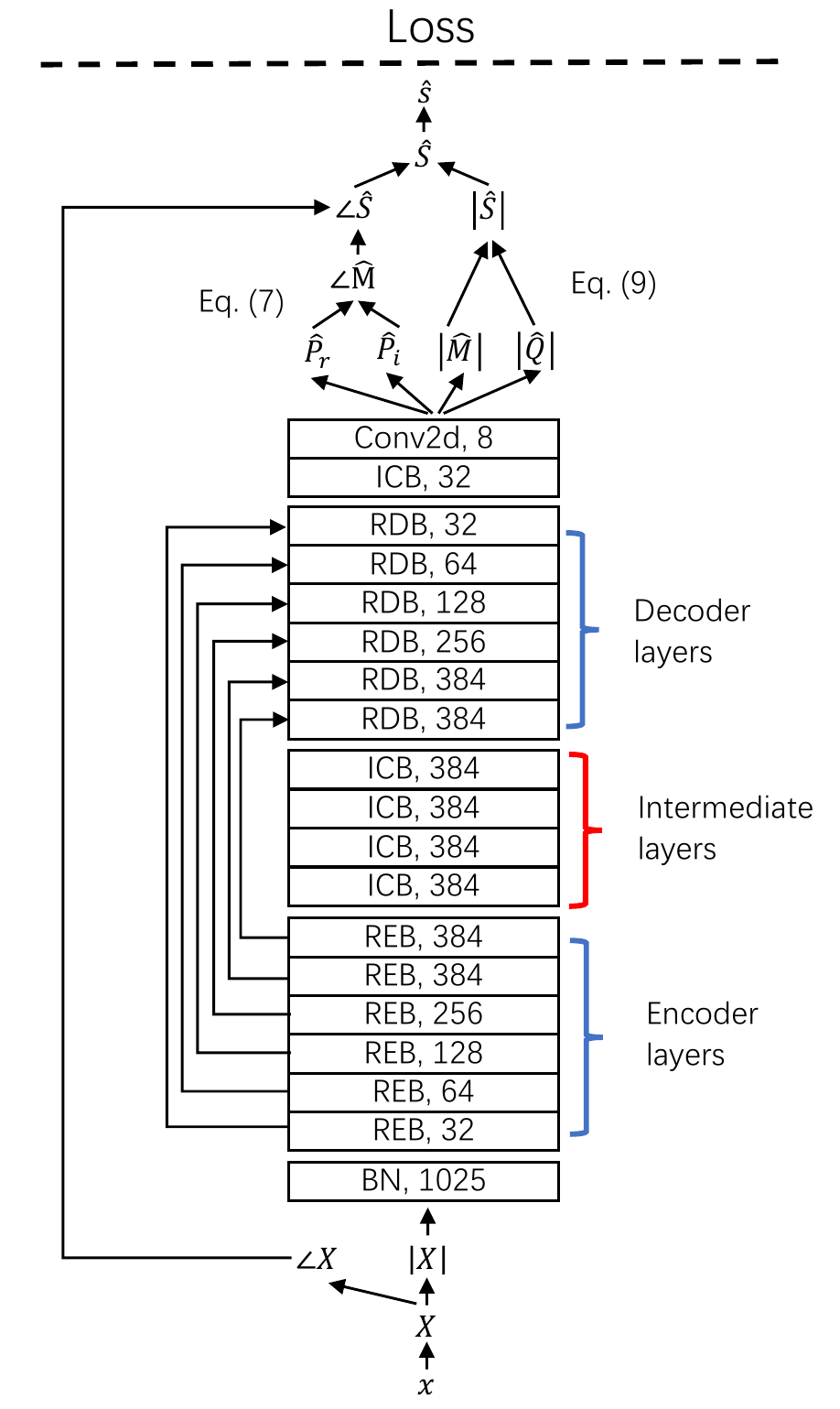}}
 \caption{The proposed MSS system with residual blocks. The details of REB, RDB, and RCB are illustrated in Figure~\ref{fig:resblocks}.}
 \label{fig:resunet}
\end{figure}

In total, the our proposed MSS system contains four outputs: $ \hat{M}_{\text{mag}} $, $ \hat{Q} $, $ \hat{P}_{\text{r}} $ and $ \hat{P}_{\text{i}} $. All of those outputs share the same backbone architecture and apply an individual linear layer to obtain their outputs. We use sigmoid non-linearity to predict $ \hat{M}_{\text{mag}} $ to ensure they have values between 0 and 1. 
Fig. \ref{fig:resunet} shows the structure of our proposed method.

\subsection{Residual UNet}\label{subsec:verydeep}
In this section, we introduce deep residual UNets with hundreds of layers for MSS, which is at least 4 times deeper than previous UNet models \cite{jansson2017singing, hennequin2020spleeter, hu2020dccrn}. 

We first introduce a baseline UNet with 33 layers. The 33-layer UNet consists of 6 encoder and 6 decoder layers. Each encoder layer consists of two convolutional layers and a downsampling layer. Each decoder layer consists of one transposed convolutional layer for upsampling and two convolutional layers. Finally, three additional convolutional layers are added after decoder layers. In total, there are 33 convolutional layers.

\begin{figure}
 \centerline{
 \includegraphics[width=\columnwidth]{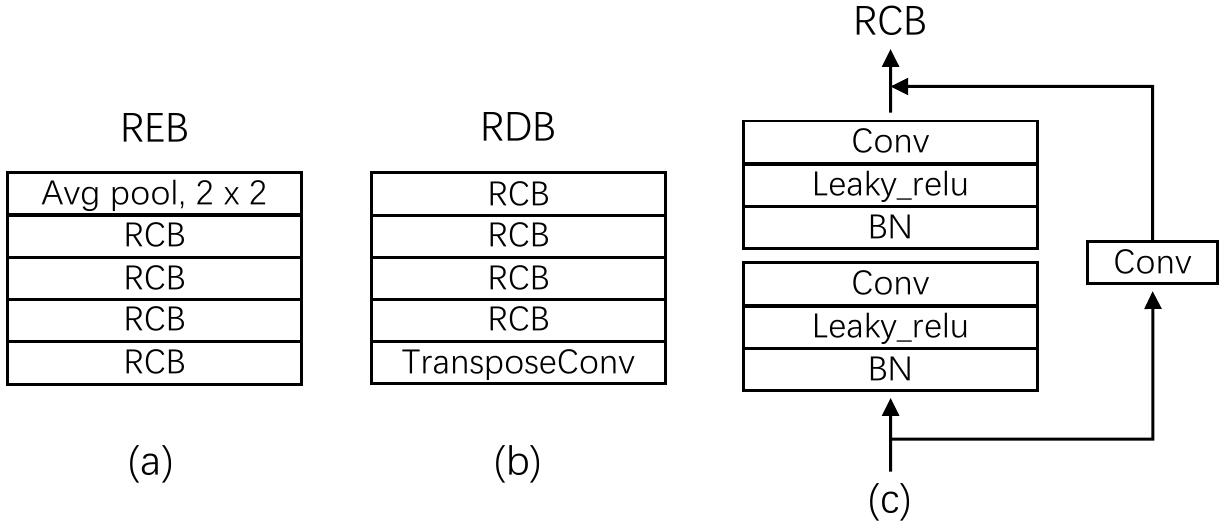}}
 \caption{(a) Residual encoder block (REB), (b) residual decoder block (RDB), (3) residual convolutional block (RCB).}
 \label{fig:resblocks}
\end{figure}

Next, we introduce a 143-layer residual UNet. In building a residual UNet with hundreds of layers, we use residual encoder blocks (REB) and residual decoder blocks (RDB) to increase its depth. Fig. \ref{fig:resunet} shows the architecture of our proposed residual UNet where we use 6 REBs and 6 RDBs. Each REB consists of 4 residual convolutional blocks (RCB) as shown in Fig. \ref{fig:resblocks} (a). Each RCB consists of of two convolutional layers with kernel sizes $ 3 \times 3 $ as shown in Fig. \ref{fig:resblocks} (c). A shortcut connection is added between the input and the output of a RCB. A batch normalization \cite{ioffe2015batch} and a leaky ReLU non-linearity \cite{maas2013rectifier} with a negative slope of 0.01 is applied before convolutional layers following the pre-act residual network configuration \cite{he2016identity}. An $ 2 \times 2 $ average pooling layer is applied after each REB to reduce the feature map size. Each REB consists of 8 convolutional layers.

The blocks in the decoder (RDBs) are symmetric to those in the encoder (REB). Each RDB consists of a transposed convolutional layer with a kernel size $ 3 \times 3 $ and stride $ 2 \times 2 $ to upsample feature maps, followed by four RCBs as shown in Fig. \ref{fig:resblocks} (b). Each RDB consists of 9 convolutional layers, including 8 convolutional layers and 1 transposed convolutional layer. To further increase the representation ability of the residual UNet, we introduce intermediate convolutional blocks (ICBs) between REBs and RDBs as shown in Fig. \ref{fig:resunet}. We use 4 ICBs, where each ICB consists of 8 convolutional layers which has the same architecture as the REB except the pooling layer.

After RDBs, an additional ICB with 8 layers and a final convolutional layer with $ J $ output channels are applied. For example, for a stereo separation task where only the magnitude of masks $ |\hat{M}| $ is used as a baseline, $ J $ is set to $2$. Similarly, if the decoupling of magnitude and phase are predicted (as in Section \ref{section:decoupling}), $ J$ is set to $6 $ (two channels of $ |\hat{M}| $, $ \hat{P}_{\text{r}} $ and $ \hat{P}_{\text{i}} $). 
In our complete system in Section \ref{section:attention}, where the combination of magnitude mask and direct magnitude prediction is used, $J$ is set to $8$ (two channels of $ |\hat{M}| $, $ |\hat{Q}| $, $ \hat{P}_{\text{r}} $ and $ \hat{P}_{\text{i}} $). 
In total, there are 143 convolutional layers in our proposed residual UNet.





\section{Experiments}
\subsection{Dataset}

We run an experiment to demonstrate the proposed method on the MUSDB18 dataset \cite{rafii2017musdb18}. The MUSDB18 dataset includes separate vocals, accompaniment, bass, drums, and other instruments. Its training/validation sets contain 100/50 full tracks, respectively. The training set is further decomposed into 86 training songs and 14 songs for development and evaluation. All songs are stereo with a sampling rate of 44.1~kHz. We release the source code of our work online.\footnote{Will be released after acceptance.}

\subsection{Data Processing}
We split audio recordings into 3-second segments. Since the proposed system is convolutional layer-based UNet, it does not require previous states to calculate current predictions, making our system to be fully parallelizable. For data augmentation, we apply \textit{mix-audio} data augmentation that is used in \cite{song2021catnet} to augment vocals, accompaniment, drums, and other instruments which randomly mix two 3-second segments from a \textit{same} source as a new 3-second segment for training. The motivation is that, the addition of two sources also belongs to that source. We do not apply mix-audio data augmentation to bass because bass are usually monophonic in a song. Then, we create mixtures $ x $ by summing segments after mix-audio augmentation from different sources. We apply short-time Fourier transform (STFT) on $ x $ with a Hann window size of 2048 and a hop size of 441 samples, corresponding to the hop size time of 10 ms. 

During training of all the proposed and baseline systems, we set batch size to 16 and apply Adam optimizer \cite{kingma2014adam}. The learning rate is set to $ 0.001 $, $ 0.0005 $, $ 0.0001 $, $ 0.0002 $, and $ 0.0005 $ for vocals, accompaniment, bass, drums and other instruments. Different learning rates are used because some sources such as drums are easier to be overfitted. Those learning rates are tuned on the validation set of the MUSDB18 dataset. Learning rates are multiplied by a factor of 0.9 after every 15,000 steps. MSS systems are trained for 300,000 steps.

\begin{table}[t]
\centering
\caption{Comparison of SDRs of previous and our proposed MSS systems.}
\label{table:previous}
\resizebox{\columnwidth}{!}{%
\begin{tabular}{*{6}{c}}
 \toprule
 & vocals & bass & drums & other & acc. \\
 \midrule
 Open-Unmix \cite{stoter2019open} & 6.32 & 5.23 & 5.73 & 4.02 & - \\
 Wave-U-Net \cite{stoller2018wave} & 3.25 & 3.21 & 4.22 & 2.25 & - \\
 Demucs \cite{defossez2019music} & 6.29 & 5.83 & 6.08 & 4.12 & - \\
 Conv-TasNet \cite{luo2019conv} & 6.81 & 5.66 & 6.08 & 4.37 & - \\
 Spleeter \cite{hennequin2020spleeter} & 6.86 & 5.51 & 6.71 & 4.55 \\
 D3Net \cite{takahashi2020d3net} & 7.24 & 5.25 & \textbf{7.01} & 4.53 & 13.52 \\
 \midrule
 ResUNetDecouple+ & \textbf{8.98} & \textbf{6.04} & 6.62 & \textbf{5.29} & \textbf{16.63} \\
 \bottomrule
\end{tabular}}
\end{table}

\subsection{Result 1: Comparison with Previous Methods}

We compare our proposed system with several systems including previous time domain and frequency domain based systems. Signal-to-Distortion Ratio~(SDR) \cite{vincent2006performance} is used as evaluation metric. The \textit{museval} toolbox \cite{stoter20182018} is used to calculate MSS metrics. 

Table \ref{table:previous} shows the SDRs of previous MSS systems as well as those of our best performing system. 
The first row shows the performance of Open-Unmix \cite{stoter2019open}, which consists of three bi-directional long short-term memory layers achieves a vocals SDR of 6.32 dB. The second row shows that the Wave-U-Net \cite{stoller2018wave} system trained in the time-domain achieve slightly lower SDRs than other time-frequency domain systems. The third to to the eighth rows show the results of Demucs \cite{defossez2019music}, Conv-TasNet \cite{luo2019conv}, Spleeter \cite{hennequin2020spleeter}, and D3Net \cite{takahashi2020d3net}. Among the compared methods, D3Net achieves the best vocals and drums SDRs of 7.24 dB and 7.01 dB respectively. The Demucs achieves the best bass SDR of 5.83 dB, and the Spleeter achieves the best other SDR of 4.55 dB in previous works. As in the last row of Table \ref{table:previous} , our proposed residual UNet with the decoupling and the combination of magnitude masks and direct prediction significantly outperforms previous methods in separating vocals, bass, other, and accompaniments. 

\subsection{Result 2: Ablation Study}

\begin{table}[t]
\centering
\caption{SDRs of the proposed systems (2nd -- 7th rows) in a comparison to the previous system, UNetPhase.}
\label{table:proposed}
\resizebox{\columnwidth}{!}{%
\begin{tabular}{*{6}{c}}
 \toprule
 & vocals & bass & drums & other & acc. \\
 \midrule
 UNetPhase \cite{choi2018phase} & 7.45 & 5.42 & 6.51 & 4.86 & 15.23 \\
 \midrule
 UNet & 7.20 & 4.79 & 5.94 & 4.49 & 14.69 \\
 UNetDecouple & 7.65 & 5.00 & 6.29 & 4.71 & 15.21 \\
 UNetDecouple+ &  7.81 & 5.28 & 6.47 & 5.00 & 15.32 \\
 \midrule
 ResUNet & 7.79 & 5.00 & 6.20 & 5.13 & 16.15 \\
 ResUNetDecouple & 8.72 & 5.71 & 6.50 & 5.20 & 16.39 \\
 ResUNetDecouple+ & \textbf{8.98} & \textbf{6.04} & \textbf{6.62} & \textbf{5.29} & \textbf{16.63} \\
 \bottomrule
\end{tabular}}
\end{table}

In this section, we show the performances of our proposed systems that partially incorporate our modification. We also compare them with the system from \cite{choi2018phase}, which we call UNetPhase. We implement a UNetPhase with 33 layers. 

In Table \ref{table:proposed}, UNet, UNetDecouple, and UNetDecouple+ are variants of a 33-layer UNet and ResUNet, ResUNetDecouple, ResUNetDecoup+ are variants of a 143-layer residual UNet. UNet and ResUNet are models with magnitude masks only, i.e., phase is not considered in the model. `Decouple` indicates that the proposed decoupling of magnitude and phase is applied. `+' indicates the further improvement of combining the magnitude masks and direct prediction as introduced in Section~\ref{section:attention}.

First, UNet, which only predicts the magnitude of masks, performed slightly worse than UNetPhase. Here, we observe the average improvement by predicting phase is 0.57~dB.

Second, we can compare the trend within the row 2-4 or the row 5-7. Both for UNet's and ResUNet's, decoupling of the magnitude and phase improves the performance -- by 0.35~dB with UNet and 0.45~dB with ResUNet on average. The `+' models shows further average improvements of 0.2~dB and 0.196~dB with UNet and ResUNet, respectively. This result indicates that combining bounded mask estimation and direct magnitude prediction can improve MSS.

Third, when the other conditions are fixed, ResUNet always outperforms UNet for all source instruments. It clearly demonstrates the effectiveness of a very deep architecture in MSS. The average improvement of ResUNet from UNet is 0.7~dB.

The results did not show a clear sign that the upper bound that we discussed in Section~\ref{sec:background} is playing a critical role in the current systems.
For example, for vocal/bass/drums/other/accompaniments, the upper bounds of cIRM (1), i.e., UNetPhase, are 19.84/17.99/19.10/18.97/26.54~dB, all of which are more than 10~dB higher than the performance of UNetPhase. Compared to UNetPhase, UNetDecouple+, which is a case of cIRM (inf), only slightly outperforms UNetPhase by 0.082~dB on average and did not perform better on bass and drums.

\section{Conclusion}
In this paper, we investigated the music source separation (MSS) task. We showed that previous MSS methods have upper bound of the performance due to a strong assumption on the magnitude of the masks. We also showed that accurate phase estimation and unbound complex ideal ratio masks (cIRMs) are important for MSS. 
Finally, we analyzed the distribution of cRIMs for MSS and showed that 22\% of cIRMs have magnitude larger than one. 
To overcome the limits, 
We proposed to decouple the estimation of magnitudes and phases. We also proposed to combine bounded magnitude masks and direct prediction methods for more flexible magnitude estimation. Finally, we proposed a very deep MSS architecture, a residual UNet with 143 layers. 
In the experiment, we showed that our proposed modifications improve the performance, achieving an SDR of 8.98~dB for vocals in MUSDB18.
In the future work, we will explore a more effective approach to design a MSS that solve the issues we analyzed better, especially, the issue of the bounded magnitude masks.

\bibliography{ISMIRtemplate}

%
%
%
%
%

\end{document}